\documentclass[aps,preprint,amsmath,amssymb]{revtex4}
\usepackage{amsfonts}
\usepackage{graphicx}
\usepackage{color}
\def\g5{\gamma_{5}}
\def\ga{\gamma}

\def\be{\begin{eqnarray}}
\def\en{\end{eqnarray}}

\def\non{\nonumber}

\def\UP{\cal U}
\def\la{\langle}
\def\ra{\rangle}
\def\BZ{\cal BZ}

\begin{document}
\title{\Large \bf Flavors and Phases in Unparticle Physics }
\date{\today}

\author{ \bf  Chuan-Hung Chen$^{1,2}$\footnote{Email:
physchen@mail.ncku.edu.tw} and Chao-Qiang
Geng$^{3}$\footnote{Email: geng@phys.nthu.edu.tw}
 }

\affiliation{ $^{1}$Department of Physics, National Cheng-Kung
University, Tainan 701, Taiwan \\
$^{2}$National Center for Theoretical Sciences, Hsinchu 300, Taiwan\\
$^{3}$Department of Physics, National Tsing-Hua University, Hsinchu
300, Taiwan
 }

\begin{abstract}
Inspired by the  recent Georgi's unparticle proposal, we study the
flavor structures of the standard model (SM) particles when they
couple to unparticles. At a very high energy scale, we introduce
$\BZ$ charges for the SM particles, which are universal for each
generation and allow $\BZ$ fields to distinguish flavor generations.
At the $\Lambda_{\UP}$ scale,  $\BZ$ operators and charges are
matched onto unparticle operators and charges, respectively. In this
scenario, we find that tree flavor changing neutral currents (FCNCs)
can be induced by the rediagonalizations of the SM fermions. As an
illustration, we employ the Fritzsch ansatz to the SM fermion mass
matrices and we find that the FCNC effects could be simplified to be
associated with the mass ratios denoted by
$\sqrt{m_{i}m_{j}/m^2_{3}}$, where $m_3$ is the mass of the heaviest
particle in each type of fermion generations and $i,\,j$ are the
flavor indices. In addition, we show that there is no new CP
violating phase  for FCNCs in down type quarks beside the unique one
in the CKM matrix. We use $\bar B_{q}\to \ell^{+} \ell^{-}$ as
examples to display the new FCNC effects. In particular, we
demonstrate that the direct CP asymmetries in the decays can be
$O(10\%)$ due to the peculiar CP conserving phase in the unparticle
propagator.

\end{abstract}
\maketitle

In the standard model (SM), it is known that
 flavor changing processes at tree level can only
be generated for charged currents mediated the
W gauge boson
 in the quark sector. These charged currents will induce flavor
changing neutral currents (FCNCs) via quantum loops. Consequently,
the most impressive features of flavor physics are the
Glashow-Iliopoulos-Maiani (GIM) mechanism \cite{GIM} and the
large top quark mass \cite{PDG06}. For instance, the former
makes the $P^0-\bar P^0$ ($P=K$ and $D$)
mixings and rare $P$
decays  naturally small while the latter leads to large $B_q-\bar
B_q$ mixings ($q=d,\, s$) as well
as the time-dependence CP asymmetry for
the decay of $B_d\to J/\Psi K_S$. Among these effects, the
most important
measured quantities are the
Cabibbo-Kobayashi-Maskawa (CKM) matrix elements \cite{CKM},
coming from the unitary matrices which diagonalize the
left-handed up and down-type quark matrices.
Although  there are no disagreements between the SM and
%{\color{red}
experiments, which might give us some clue as to what may lie beyond
the SM,
it is important to keep searching for any discrepancies.
In particular,
the next generation of
flavor factories such as SuperKEKB \cite{SuperKEKB}
and LHCb \cite{LHCb} with design luminosities of
$5\times 10^{35}$ and $5\times 10^{32}\;cm^{-2}s^{-1}$, respectively,
%\
may provide some hints for new flavor effects.
Thus, theoretically it should be interesting to  explore the
possible new phenomena related to flavor physics
\cite{newphys1,newphys2}.

Recently, Georgi has proposed that an invisible sector dictated by
the scale invariant
may weakly couple to the particles of the SM
\cite{Georgi1, Georgi2}.
Since the scale invariant stuff
cannot have a definite mass unless it is zero,
it should be made of {\it
unparticles}
\cite{Georgi1} as the SM particles have definite masses. In terms of
the two-point function with the scale invariance,
it is found that
the unparticle with the scaling dimension $d_{\UP}$ behaves like a
non-integral number $d_{\UP}$ of invisible particles \cite{Georgi1}.
Consequently, the unparticle physics phenomenology has been
extensively explored in Refs.
\cite{Georgi1,Georgi2,CKY,LZ,CH12,Ding_Yan,Y_Liao,Aliev,Catterall,
Li,Lu,Steph,Fox,Greiner,Davou,CGM,XG,MR,Zhou,FFRS,Bander,Rizzo,GN,
Zwicky,Kikuchi,M_Giri,Huang,Krasnikov,Lenz,C_Ghosh,Zhang,Nakayama,
Deshpande,DEQ,Neubert,Hannestad,Das,Bhattacharyya,Majumdar,Alan-Pak}.
%TC,Luo,CH,Yan,Liao,Aliev,Catterall,Li,Lu,Steph,Fox,
%Greiner,Davou,Choudhury,XG,Aliev2,MR,Zhou,DY,CH2,
%Liao:2007ic,Bander:2007nd,Rizzo:2007xr,Cheung:2007ap,Goldberg:2007tt,
%Chen:2007zy,Zwicky:2007vv,Kikuchi:2007qd,Mohanta:2007ad,
%Huang:2007ax,Krasnikov:2007fs,Lenz:2007nj,Choudhury:2007cq,Zhang:2007ih,
%Li:2007kj,Nakayama:2007qu,Deshpande:2007jy,Mohanta:2007uu,
%Delgado:2007dx,Neubert:2007kh,Luo:2007me,Hannestad:2007ys,
%Deshpande:2007mf,Das:2007nu,Bhattacharyya:2007pi,Liao:2007fv}.
%In particular, s
Some illustrative examples such as $t\to u+ \UP$ and $e^{+} e^{-}\to
\mu^{+} \mu^{-}$  have been given to display the unparticle
properties. It is also suggested that the unparticle production in
high energy colliders might be detected by searching for the missing
energy and momentum distributions \cite{Georgi1,Georgi2}.
Nevertheless, flavor factories with high luminosities mentioned
above should also provide good environments to search for
unparticles via their virtual effects.

Besides the Lorentz structure,
so far there is no rule to
govern the interactions between the SM particles and unparticles.
The flavor
physics associated with unparticles is quite arbitrary, {\it i.e.},
the couplings could be flavor conserving or changing.
Moreover,
 there is no any correlation in the
transitions among three generations
for flavor changing processes.
In this
note, we
will study the possible flavor structures for the SM
particles when they couple to unparticles. Since the gauge
structure of unparticles involves more theoretical uncertainties,
we only pay attention to the interactions with
the charged fermion sectors.
We will not discuss the neutrino sector
because the nature of neutrino flavors is still unclear.

We start from the scheme proposed in Ref.~\cite{Georgi1}. For the
system with the scale invariance \cite{Georgi1} there exist so-called
Banks-Zaks ($\cal BZ$) fields that have a nontrivial infrared fixed
point at a very high energy scale \cite{BZ}.
 Above the electroweak scale, since
all SM particles are massless,
we cannot tell the
differences between down-type quarks or up-type quarks.
In the SM, we have $SU(3)_D\times SU(3)_U\times SU(3)_{Q}$ flavor
symmetries \cite{MM}, where $D$ and $U$ denote the singlet states
for down and up-type quarks, respectively, while $Q$ stands for the
quark doublet. Therefore, if $\BZ$ fields are flavor blind,
plausibly new flavor mixing effects cannot be generated for vector
and axial-vector currents after the electroweak symmetry breaking
(EWSB). It is worthy to mention that scalar-type couplings,
illustrated by
%$\bar d \,P_Ld{\cal O}$
$\bar d \,d{\cal O}$ and $\bar d \,\gamma_5d{\cal O}$
in weak eigenstates,
could basically produce FCNCs after the spontaneous symmetry
breaking.
%{\color{red}
For example, $\bar d_{i} \left(V^D_R
V^{D^\dagger}_L\right)_{ij} P_Ld_{j} {\cal O}$ would be induced for the coupling
of $\bar d \,P_Ld{\cal O}$ after the EWSB, where
$V^{D}_{R,L}$ are unitary matrices to diagonalize the down-type
quark Yukawa matrix. Note that since $\bar d\Gamma d$ ($\Gamma=1,\gamma_5$) have
to be  $SU(2)_L$ singlets, the $d$-quark has to be either left-handed or right-handed before
the EWSB as it should be.
%}
For the convention of $V^{U}_L=1$, $V^{D}_{L}$
is just the CKM matrix. Immediately, we suffer a serious problem
from  the $K^0-\bar K^0$ mixing due to the coupling for $\bar d s
{\cal O}$ being associated with ($V^{D}_{R12}-\lambda$) where
$\lambda$ is the Wolfenstein parameter \cite{Wolfenstein}.
To avoid the large FCNC problem,
one can set the Yukawa matrix be hermitian so that
$V^{D}_{R}=V^{D}_{L}$. As a result, the FCNCs at tree level via
scalar-type interactions are removed. In any event, despite the
property of Yukawa matrices, to get natural small FCNCs at tree level
for scalar and vector-type interactions, we need some internal
degrees of freedom for fermions that could differentiate flavors
by the scale invariant stuff.

In order to reveal the new flavor mixing effects due to the
involvement of unparticles, we assume that the SM particles carry
some kind of $\BZ$ charges so that $\BZ$ fields could distinguish
flavor species. In terms of the prescription in Ref.~\cite{Georgi1},
the interactions between $\BZ$  and SM fields are
given by
\be
\frac{g_{\BZ}}{M^{k}_{\UP}} \bar F {\bf Q^{\BZ}} \Gamma F {\cal
O_{BZ}} \label{eq:BZint}
\en
 where $M_{\UP}$ is the high energy mass scale of the messenger,
$g_{\BZ}$ is a free parameter, $F^{T}=(f_1, f_2, f_3)$ denote the
3-generation of fermions in the SM, $\rm dia{\bf
Q}^{\BZ}=(Q^{\BZ}_1, Q^{\BZ}_2, Q^{\cal BZ}_3)$ are the
corresponding ${\BZ}$ charges, $\Gamma$ is the possible Dirac matrix
and ${\cal O}_{\BZ}$ is the operator composed of $\BZ$ fields. We
note that although $Q^{\BZ}_i$ are different for each generation,
the interactions are still flavor conserved. To simplify our
discussion, we regard that all fermions in each generation have the
same $\BZ$ charge at the high energy scale and we assume that the
interactions with the $\BZ$ fields are  invariant under parity.
Subsequently, with the dimensional transmutation at the $\Lambda_{\UP}$
scale, the $\BZ$ operators in Eq.~(\ref{eq:BZint}) will match onto
unparticle operators.
The effective interactions are
obtained to be
\be
C^{F}_{\UP}\frac{ \Lambda^{d_{\BZ}}_{\UP} }{M^k_{\UP}
\Lambda^{d_{\UP}}_{\UP}} \bar F {\bf Q^{\UP}} \Gamma F {\cal
O}_{\UP}\,, \label{eq:UPint}
\en
where $C^{F}_{\UP}$ is a Wilson-like coefficient function and
$d_{\BZ(\UP)}$ is the scaling dimension of the $\BZ$ (unparticle)
operator. Here the unparticle operators have been set to be hermitian
\cite{Georgi2}. In principle, ${\bf Q}^{\UP}$ could be related to
${\bf Q}^{\BZ}$ by a complicated matching procedure. However, at the
current stage, it is impossible to give any explicit calculations
for the matching.
With the property of the diagonal ${\bf Q^{\BZ}}$ matrix, we
know that ${\bf Q}^{\UP}$ should be also a diagonal one,
parametrized by ${\rm dia}{\bf Q}^{\UP}=(Q^{\UP}_{1} ,\, Q^{\UP}_{2}
,\, Q^{\UP}_3)$. Hence, below the $\Lambda_{\UP}$ scale, ${\bf Q}^{\UP}$
could be regarded as
unparticle charges carried by the SM fermions to
distinguish the flavors by the unparticle stuff.

When the energy scale goes down below the EWSB scale,
described by the vacuum expectation value (VEV) of the
Higgs field $\la
H \ra=v/\sqrt{2}$, the flavor symmetry will be broken by the Yukawa
terms and
the charged fermions become massive. Afterward,
the weak eigenstates of the fermions appearing in Eq.~(\ref{eq:UPint})
need to be transformed to the physical eigenstates by proper unitary
transformations. Hence, Eq.~(\ref{eq:UPint}) is found to be
\be
{\cal L}_{\UP}&=&\frac{ C^F_S }{\Lambda^{d_{\UP}-1}_{\UP}}
\left(\bar F V^{F}_{R} {\bf Q^{\UP}} V^{F^\dagger}_{L} P_{L} F +
h.c.\right) {\cal O}_{\UP} \non\\
&+& \frac{ 1 }{\Lambda^{d_{\UP}-1}_{\UP}} \left(\bar F V^{F}_{L}
{\bf Q^{\UP}}  V^{F^\dagger}_{L} \ga_{\mu} P_{L} F+ L\to R\right)
\left( C^F_V {\cal O}^{\mu}_{\UP} +\frac{ C^F_{VS}}{\Lambda_{\UP}}
\partial^{\mu} {\cal O}_{\UP} \right) +\ldots \,,
\label{eq:UPint_phy}
\en
where we have redefined the coefficient functions
to be dimensionless free parameters
denoted by $C^F_{S}$, $C^F_{V}$ and $C^F_{VS}$,
respectively.
 In Eq. (\ref{eq:UPint_phy}), the power
of $\Lambda_{\UP}$ is taken to fit the dimension of the
effective Lagrangian in four-dimension spacetime and
the explicit terms
represent the main FCNC effects.
Note that we have separated the interactions in terms of the
fermion chirality. In addition, $V^{F}_{R, L}$ are the unitary
matrices to diagonalize the Yukawa matrix of F-type fermions,
where F could be up and down-type quarks and charged leptons.
According to Eq.~(\ref{eq:UPint_phy}), in general we have two types
of sources for new FCNCs, {\it i.e.}, $V^{F}_{R} {\bf Q}^{\UP}
V^{F^\dagger}_{L}$ and $V^{F}_{L(R)} {\bf Q}^{\UP}
V^{F^\dagger}_{L(R)}$. As known, the determination of flavor mixing
matrices $V^{F}_{L, R}$ is governed by the detailed patterns of the
mass matrices.
For convenience, we just focus on the quark sector.
It has been known that the CKM matrix, defined by $V^{U}_{L}
V^{D^\dagger}_{L}$, is approximately an unity matrix.
This indicates
that the quark mass matrices
are very likely aligned and have the relationship of ${\cal
M}_{D}={\cal M}_{U} + \Delta(\lambda^2)$ with ${\cal
M}_{U(D)}=M_{U(D)}/m_{t(b)}$ \cite{qm1,qm2,qm3}. In Ref.~\cite{qm3},
it showed that the Fritzsch quark mass matrices, given by
\cite{Fritzsch,qm2}
\be
M_{F}=R_{F} \bar M_{F} H_{F}\ { \rm with}\ \bar M_{F}=\left(
        \begin{array}{ccc}
          0 & A_{F} & 0 \\
          A_{F} & 0 & B_{F} \\
          0 & B_{F} & C_{F}  \\
        \end{array}
      \right) \label{eq:mass}
\en
where $R_{F}$ and $H_{F}$ are diagonal phase matrices, could lead to
reasonable structures for the mixing angles and CP violating phase
in the CKM matrix just in terms of the quark masses.
%{\color{red}
 From the hierarchy $m_{u(d)}\ll m_{c(s)}\ll m_{t(b)}$, it is found
that the interesting equalities \cite{qm3}
\be
\sqrt{m_d/m_s}-\sqrt{m_u/m_c} \approx V_{us}\,,\non\\
\sqrt{m_s/m_b}-\sqrt{m_c/m_t} \approx V_{cb}
\en
are satisfied.
%.
 Although the extensions of the
Fritzsch ansatz could have more degrees of freedom to fit the
experimental data \cite{MN}, however, since our goal of this study
is to explore the flavor structure affected by unparticles, we will
take the simplest version of the Fritzsch ansatz in Eq.
(\ref{eq:mass}) as our working base. In addition, we have checked
that due to the character of mass hierarchy, the extensions of
Eq.~(\ref{eq:mass}) do not change
our following results.

Since the SM has been extended to include new flavor interactions,
we have to be careful to use the phase convention because the
rotated phases will flow to Eq.~(\ref{eq:UPint_phy}). To avoid the
phase ambiguity, we should start from the flavor basis in
Eqs.~(\ref{eq:UPint}) and (\ref{eq:mass}). At first, we rotate away
$R_{U}$ and $H_U$ from $M_{U}$ by redefining the phases of the
up-type quarks.
%{\color{red}
In order to make the weak charged
currents to be invariant under this transformation, left-handed down
quarks should make the transformation $d_{L} \to  H_{U} d_{L}$
simultaneously. Then, the interactions in Eq.~(\ref{eq:UPint}) for
up and down quarks to the scalar unparticle become
 \be
 \nonumber
  \bar u_R  {\bf Q}^{\UP}[R_{U} H^{\dagger}_{U}] u_{L} {\cal O}_{\UP}, \ \ \ \bar d_{R}
  {\bf Q}^{\UP}
  H_{U}^{\dagger} d_{L} {\cal O}_{\UP}\,.
 \en
Due to ${\bf Q}^{\UP}$, $R_F$ and $H_{F}$ being all diagonal matrices, the
phase redefinition will not influence the vector current
interactions.
%}
Since $\bar M_{F}$ is a real and symmetric matrix, it
can be diagonalized by a orthogonal matrix $O_{F}$ such that
 $\bar M_{F}^{dia}= O_F \bar M_{F} O^{T}_{F}$.
Accordingly, we obtain $V^{U}_{L}=V^{U}_R=O_{U}$, $V^{D}_{L}=O_D
H_{D} H^{\dagger}_{U}$ , $V^{D}_{R}=O_{D} R_{D}$ and $V_{CKM}=O_{U}
H_{U} H^{\dagger}_{D} O^{T}_D$. Then, the flavor structures
 could be expressed by
\be
\bar F \left[O_{F} {\bf Q}^{\UP} R_{F} H^{\dagger}_{F} O^{T}_{F}
\right] P_L
F + h.c.\,, \non\\
\bar F \left[O_{F} {\bf Q}^{\UP}  O^{T}_{F} \right] \ga_{\mu} P_L F
+ (P_R\to P_L)\,. \label{eq:FCNC_ph1}
\en
We note that the phases in $R_{F}$ and $H_F$ appear only in the
scalar-type interactions. With $tr \bar M_{F}$, $tr\bar M^2_F$ and
$det \bar M_{F}$ and the convention of ${\rm dia}\bar
M^{dia}_{F}=(m_1, -m_2, m_3)$ where $m_{1,2,3}$ denote $m_{u, c,
t[d, s, b]}$ and $F=U[D]$,  we find that $A_{F}\approx \sqrt{m_1
m_2}$, $B_{F}\approx \sqrt{m_2 m_3}$ and $C\approx m_3$. As a
result, the orthogonal matrix could be obtained as \cite{qm3}
\be
O_{F}&\approx& \left(
           \begin{array}{ccc}
             1-m_1/2m_2 & \sqrt{m_1/m_2} &  -\sqrt{m_1/m_3}  \\
           -\sqrt{m_1/m_2}  & 1-m_1/2m_2-m_2/2m_3 & -\sqrt{m_2/m_3}  \\
             \sqrt{m_1/m_3} m_2/m_3 & \sqrt{m_2/m_3} & 1-m_2/2m_3 \\
           \end{array}
         \right)\,. \label{eq:omatrix}
\en Since the CKM matrix expressed by $V_{CKM}=O_{U} H_{U}
H^{\dagger}_{D} O^{T}_D$ in general has six phases, we can redefine the
phases in up and down-type quarks again so that $V_{CKM}=X O_{U}
H_{U} H^{\dagger}_{D} O^{T}_D Y^{\dagger}$ satisfies one single CKM
phase convention \cite{PDG06}. With the new phases in ${\rm dia
X}=e^{i(\alpha-\beta)}(-i, 1, 1)$, ${\rm dia}Y=e^{i\alpha}(-1, 1,
1)$ and ${\rm dia}H_{U}H^{\dagger}_{D}=e^{i\beta}( - i, 1, 1)$,
Eq.~(\ref{eq:FCNC_ph1}) becomes
\be &&\bar F  Z_{F} \left[O_{F} {\bf
Q}^{\UP} R_{F} H^{\dagger}_{F} O^{T}_{F} \right]
Z^{\dagger}_{F} P_{L} F + h.c.\,,\non\\
&&\bar F Z_{F} \left[O_{F} {\bf Q}^{\UP}  O^{T}_{F} \right]
Z^{\dagger}_{F} \ga_{\mu} F  \,, \label{eq:FCNC_ph2}
\en
where ${\rm dia}Z_{U}=(-i, 1, 1)$, ${\rm dia}Z_{D}=(-1,1,1)$ and the
vector-type interactions are parity conserved.
We note that the
flavor structures in Eq.~(\ref{eq:FCNC_ph2}) have some restrictions
on ${\bf Q}^{\UP}$. To see the problem clearly,  we decompose the
flavor changing effects to be
\be
\left(O_{F} {\bf Q}^{\UP} O^{T}_{F}\right)_{ij} &=& Q_{1}
\left[\delta_{ij} + (r_{21} - 1 ) O_{F i2} O_{Fj2} + (r_{31} -1)
O_{F i3} O_{F j3}\right] \label{eq:decomp}\non
\en
with $r_{ij}=Q^{\UP}_{i}/Q^{\UP}_{j}$. Since all  phase matrices are
in diagonal forms, an analysis on $O_{F} {\bf Q}^{\UP} O^{T}_{F}$
will not lose the generality. Using the elements in
Eq.~(\ref{eq:omatrix}), the possible flavor changing effects are
explicitly given by
\be
\left(O_{F} {\bf Q}^{\UP} O^{T}_{F}\right)_{12} &=& Q^{\UP}_1 \left[
(r_{21}-1) \sqrt{\frac{m_1}{m_2}} + (r_{31}-1)\sqrt{\frac{m_1
m_2}{m^2_{3}}}
\right]\,,\non\\
\left(O_{F} {\bf Q}^{\UP} O^{T}_{F}\right)_{13} &=&
Q^{\UP}_1(r_{21}-r_{31}) \sqrt{\frac{m_1}{m_3}}\,,\non\\
\left(O_{F} {\bf Q}^{\UP} O^{T}_{F}\right)_{23} &=& Q^{\UP}_1
(r_{21}-r_{31}) \sqrt{\frac{m_2}{m_3}}\,. \label{eq:FCNC_ph3}
\en
Phenomenologically, $\left(O_{U} {\bf Q}^{\UP}
O^{T}_{U}\right)_{12}$ and $\left(O_{D} {\bf Q}^{\UP}
O^{T}_{D}\right)_{12, 13, 23}$ are dictated by $D^0-\bar D^0$,
$K^0-\bar K^0$, $B_d-\bar B_d$ and $B_s-\bar B_s$ mixings,
respectively. From Eq.~(\ref{eq:FCNC_ph3}), one can easily see that
if $r_{21}-1\sim O(\lambda)$, a strict constraint on $Q^{\UP}_{1}$
is inevitable due to $\sqrt{m_d/m_s}\sim \lambda$. If $r_{21}$ and
$r_{31}$ are in the same order of magnitude, it will make the FCNC
effects involving the third generation be less interesting.
Motivated by the successful SM results for pseudoscalar meson
oscillations in the down-type quark systems, where the related CKM
matrix elements for $\Delta m_{K}$, $\Delta m_{B}$ and $\Delta
m_{B_s}$ have the ratios $ \lambda^3 : \lambda :1$,  we find that
$|r_{21}-1|\sim O(\lambda^2)$ in $(O_{D} {\bf Q}^{\UP}
O^T_{D})_{12}$ should be satisfied, {\it i.e.}, $Q^{\UP}_{1}\sim
Q^{\UP}_{2}+O(\lambda^2)$. In addition, the sign and the specific
magnitude should be chosen to somewhat cancel out the second term of
the first line in Eq.~(\ref{eq:FCNC_ph3}). With this scheme, we then
have
\be
(r_{21}-1)\sqrt{\frac{m_d m_b}{m^2_s}} - \sqrt{\frac{m_d}{m_b}} \sim
O\left(\sqrt{\frac{m_d m_s}{m^2_b}}\right)\,,
\en
which is needed for the phenomenological reason.

With the experimental data and
Fritzsch ansatz, we have
obtained the FCNC effects from the couplings of quarks and unparticles.
 According to the results in Eq.~(\ref{eq:FCNC_ph2}), we
highlight some interesting characters as follows:

$\bullet$ If the phase matrices $R_F$ and $H_{F}$ are independent,
from Eq.~(\ref{eq:FCNC_ph2}) we find that only scalar-type FCNCs
could have different couplings for different chiralities.
Even there are some new CP violating phases in ${\cal N}_{F}\equiv
Z_{F} \left[ O_{F} {\bf Q}^{\UP} R_{F} H^{\dagger}_{F} O^{T}_{F}
\right] Z^{\dagger}_{F}$, due to
%{\color {red}
${\cal N}^{\dagger}_F={\cal N}^{*}_F$, we see that the scalar-type
interactions are in fact associated with $\bar F \left( Re{\cal
N}_F- i Im{\cal N}_F \ga_5 \right) F$.
Thus, there are no new physical CP violating effects
 unless the processes
involve $Re{\cal N}_F\cdot Im{\cal N}_F$.
It is also true for cases with the vector current couplings.

$\bullet$ If $R_F=H_F$, from Eq.~(\ref{eq:mass}) we can easily find
that the corresponding mass matrices are hermitian. The FCNC effects
are all related to $Z_{F} \left[O_{F} {\bf Q}^{\UP} O^{T}_{F}
\right] Z^{\dagger}_{F}$ which is also hermitian. As a result, in
terms of the quark currents, the couplings of fermions and
unparticles are parity-even and no new CP phase is induced for down
type quarks in this case. It should be worthy to mention that the
hermitian mass matrices could be naturally realized in gauge models
such as left-right symmetric models \cite{hermitian_y}. The
hermiticity could help us to solve the CP problem in models with
supersymmetry (SUSY) \cite{ABKL}
 and it has  an
important implication on  CP violation in Hyperon decays \cite{Chen_PLB521}.

$\bullet$ From Eq.~(\ref{eq:FCNC_ph3}), we find that
$\Delta m_{B_s}/\Delta m_{B_d}\approx m_d/m_s \sim
\lambda^2 $, which
is consistent with the
experimental data \cite{PDG06}.

$\bullet$ Since the masses of the charged leptons also have the mass
hierarchy $m_{e}\ll m_{\mu}\ll m_{\tau}$, if we take the same phase
convention,
Eqs.~(\ref{eq:FCNC_ph2}) and
(\ref{eq:FCNC_ph3}) should be straightforwardly extended
to the charged lepton sector.

To illustrate the peculiar phenomena in Eq.~(\ref{eq:FCNC_ph2})
associated with unparticles,
 we take $\bar B_{q}\to \ell^{+} \ell^{-}$
as examples. For simplicity, we adopt scalar-type interactions
as the representative.
The effective
interactions are
\be
{\cal L}_{\UP}= \frac{ 1}{\Lambda^{d_{\UP}-1}_{\UP}} \bar q \left(
{\cal N}_{ qb} P_{L} + {\cal N}^*_{ qb} P_{R} \right) b +
\frac{1}{\Lambda^{d_{\UP}-1}_{\UP}} \bar \ell \left( {\cal N}_{\ell
\ell } P_{L} + {\cal N}^*_{\ell \ell} P_{R} \right) \ell \non
\en
with
\be
{\cal N}_{qb}&=& \sqrt{\frac{m_q}{m_b}} \left( \bar Q^{\UP}_{3}
e^{i\theta_3} - \bar Q^{\UP}_{2} e^{i\theta_2} \right)\,,\non\\
{\cal N}_{\ell \ell} &=& \bar Q^{\UP}_{\ell}
e^{i\theta_{\ell}}\,.\non
\en
Since the coefficient functions are always associated with
$\UP$-charges, we define
$\bar
Q^{\UP}_{i}=C^{D}_{S} Q^{\UP}_{i}$ and $\bar Q^{\UP}_{\ell}=
C^{L}_{S} Q^{\UP}_{i}$. With the propagator of the scalar unparticle
operator proposed in Refs.~\cite{Georgi1,Georgi2}, given by
\be
\int e^{iq x}\la 0| T{\cal O}_{\UP}(x) {\cal O}_{\UP}(0)|0\ra &=& i
\frac{A_{d_{\UP}}}{2\sin d_{\UP}\pi}
\frac{e^{-i\phi_{\UP}}}{\left(q^{2}\right)^{2-d_{\UP}}}\,,\non\\
A_{d_{\UP}}&=& \frac{16 \pi^{5/2}}{(2\pi)^{2d_{\UP}}}
\frac{\Gamma(d_{\UP}+1/2)}{\Gamma(d_{\UP}-1) \Gamma(2
d_{\UP})}\,,\non\\
\phi_{\UP}&=&(d_{\UP}-2)\pi\,,\non
\en
the decay amplitudes for $\bar B_q\to \ell^{+} \ell^{-}$ by
due to unparticles are expressed by
\be
A(\bar B_{q}\to \ell^{+} \ell^{-})&=&i \frac{f_{B_q}}{m_{B_q}}
\left(\frac{m^2_{B_q}}{\Lambda^{2}_{\UP}}\right)^{d_{\UP}-1}
 \frac{A_{d_{\UP}}}{2\sin
d_{\UP}\pi}e^{-i\phi_{\UP} } \non\\
&\times& Im{\cal N}_{qb} \left[ Re{\cal N}_{\ell\ell} \bar \ell \;
\ell - i Im {\cal N}_{\ell\ell} \bar \ell \ga_5 \ell\right]\,.\non
\en
We note that  $\phi_{\UP}$ is a CP conserving phase
%{\color{red}
\cite{Georgi2,CKY,CH12}.
Combining with the SM contributions, the corresponding branching
ratios (BRs) are
\be
\frac{1}{m^2_{\ell}}{\cal B}( B_{q}\to \ell^{+} \ell^{-})&=&
\kappa_{B_q} \left[\left|\Sigma^{B_q}_{SM} e^{i\beta_{q}} +
\Sigma^{B_q}_{\UP}e^{-i\phi_{\UP}}\right|^2 + \left|\bar
\Sigma^{B_q}_{\UP}\right|^2\right]\,, \label{eq:BrBll}
\en
where the angle $\beta_{q}$ is
from $V_{tq}=|V_{tq}|e^{-i\beta_{q}}$
with
\be
\beta_{d(s)}&=&\beta (0)\,,\non\\
\kappa_{B_q}&=& \frac{1}{m^2_{\tau}} \frac{ \alpha^2_{\rm em} {\cal
B}(B^+\to \tau^{+} \nu_{\tau} )}{\pi^2\sin^4\theta_W}
\frac{m_{B_q} f^2_{B_q}}{m_{B^+} f^2_{B^+}}\frac{\tau_{B_{q}}}{ \tau_{B^+}}\,, \non \\
\Sigma^{B_q}_{SM}&=& \frac{|V_{tb} V^{*}_{tq}|}{|V_{ub}|} Y(m^2_t/m^2_{W})\,,\non\\
\Sigma^{B_q}_{\UP}&=& \frac{8\pi\sin^2\theta_{W}}{g^2\alpha_{\rm
em}|V_{ub}|}\frac{m^2_{W}}{m_{\ell} m_{B_q}}
\frac{A_{d_{\UP}}}{2\sin d_{\UP}\pi}\left(
\frac{m^2_{B_q}}{\Lambda^2_{\UP}}\right)^{d_{\UP}-1} Im{\cal N}_{qb}
Im{\cal N}_{\ell\ell}\,,\non\\
\bar \Sigma^{B_q}_{\UP}&=&\Sigma^{B_q}_{\UP}Re{\cal
N}_{\ell\ell}/Im{\cal N}_{\ell\ell}\,.\non
\en
%.
 Due to $m_W\ll m_t$, the
function of $Y(m^2_t/m^2_W)$ can be simplified to $Y(x_t)= 0.315
x^{0.78}_ t$ \cite{BBL}. Here, we have used the measured $B^-\to
\tau \bar \nu_{\tau}$ decay to remove the uncertainties from $f_{B}$
and $|V_{tq}|$. Besides the BRs,
from Eq.~(\ref{eq:BrBll}) we
can also study the direct CP asymmetries (CPAs) in the two-body
exclusive B decays, defined by
\be
A_{CP}(B_{q}\to \ell^{+}\ell^{-})&=& \frac{{\cal B}(\bar B_q \to
\ell^{+} \ell^{-}) - {\cal B}( B_q \to \ell^{+} \ell^{-})}{{\cal
B}(\bar B_q \to \ell^{+} \ell^{-}) + {\cal B}( B_q \to \ell^{+}
\ell^{-})}\,.
\en
It is known that in a process
the direct CPA needs CP conserving and unrotated CP
violating phases simultaneously. Since the
unparticle stuff provides a CP-conserved phase, if the
process carries a physical CP violating phase, a nonvanishing CPA is
expected. In $B_{q}\to\ell^{+} \ell^{-}$,
the new free
parameters are $d_{\UP}$, $\Lambda_{\UP}$ and ${\cal N}_{qb}$
(${\cal N}_{\ell \ell}$),
%.
which can be constrained by $\Delta m_{B_q}$ ($\Delta a_{\ell}$)
of the $B_{q}-\bar B_{q}$ mixings (lepton anomalous magnetic dipole moments).
Explicitly, we find that
\be
\Delta m_{B_q}&=& 2 Re\la B_q| H_{\UP}(|\Delta B|=2) | \bar B_{q}\ra\non\\
&=& \frac{1}{6} \frac{f^2_{B_q}}{m_{B_q}} \frac{A_{d_{\UP}}}{2\sin
d_{\UP}\pi}
\left(\frac{m^2_{B_q}}{\Lambda^2_{\UP}}\right)^{d_{\UP}-1} \cos
\phi_{\UP} \left[ (Re{\cal N}_{qb})^2 + 6 (Im {\cal
N}_{qb})^2\right]\,, \non\\
\Delta a_{\ell}&=& -\frac{1}{4\pi} \frac{A_{d_{\UP}}}{2\sin
d_{\UP}\pi} \left(\frac{m^2_{\ell}}{\Lambda^2_{\UP}}
\right)^{d_{\UP}-1} \non\\
&\times&\left[ Re({\cal N}_{\ell\ell})^2
\frac{\Gamma(2-d_{\UP})\Gamma(2d_{\UP}-1)}{\Gamma(d_{\UP}+1)} +
|{\cal N}_{\ell\ell}|^2
\frac{\Gamma(3-d_{\UP})\Gamma(2d_{\UP}-1)}{\Gamma(d_{\UP}+2)}\right]\,.
\en

\begin{figure}[hpbt]
\includegraphics*[width=4.5 in]{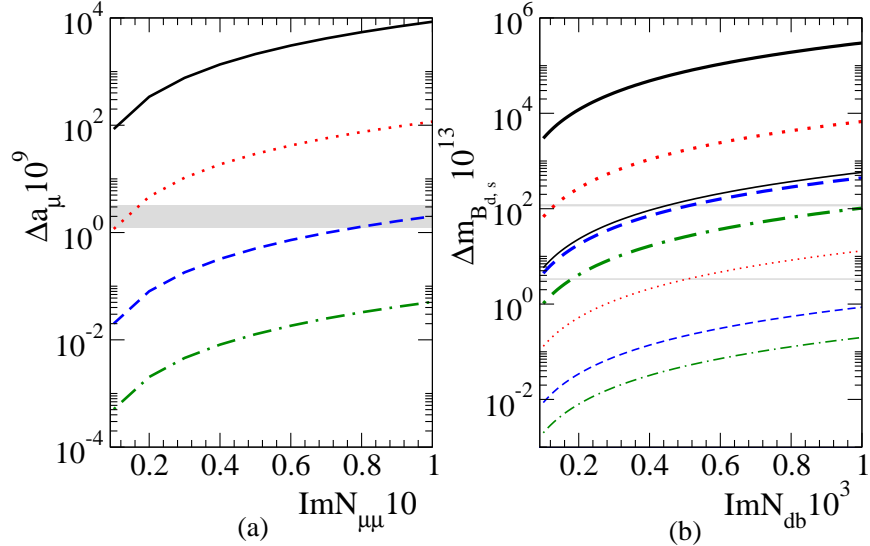}
\caption{(a) $\Delta
a_{\mu}$
and (b) $\Delta m_{B_{d, s}}$
as a function of $Im {\cal
N}_{\mu\mu}$ and $Im{\cal N}_{db}$, where the solid, dotted, dashed
and dash-dotted lines stand for $d_{\UP}=1.2$, $1.4$, $1.6$ and
$1.8$, respectively. The bands in (a)
denote the world
average with $1\sigma$ errors and
the lower (upper) band and
 thin (thick) lines in (b) are for the $B_{d(s)}-\bar{B}_{d(s)}$
 mixing. }
 \label{fig:mB-g_2}
\end{figure}

\begin{figure}[hpbt]
\includegraphics*[width=4.5 in]{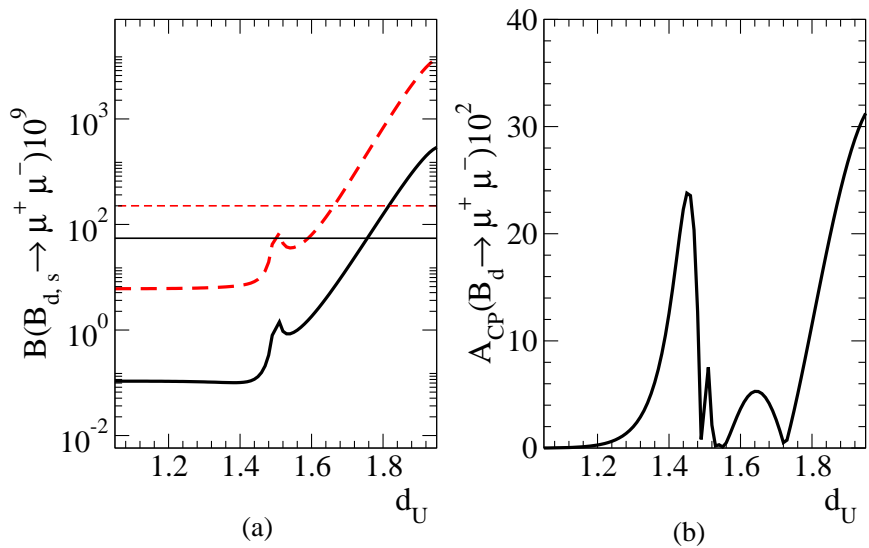}
\caption{(a) BRs
for $\bar B_{d}\to \mu^{+}
\mu^{-}$ [solid] and $\bar B_{s}\to \mu^{+} \mu^{-}$ [dashed] and
(b) CPA
%(in units of $10^{-2}$)
for $\bar B_{d}\to \mu^+ \mu^-$ as
functions of the scale dimension $d_{\UP}$, where the horizontal lines
correspond to the current experimental upper limits. }
 \label{fig:BR-CP}
\end{figure}

To estimate the numerical values, we take $|V_{ub}|=4.3\times
10^{-3} $, $V_{td}=7.4\times 10^{-3} e^{-i\beta}$ with
$\beta=25^{\circ}$, $V_{ts}=-0.041$, $m_{B_{d(s)}}=5.28$  $(5.37)$ GeV,
$f_{B_{d(s)}}=0.2$ $(0.22)$ GeV and $\sin^2\theta_{W}\approx 0.234$.
For the mixing parameters of $\Delta m_{B_d}$ and $\Delta m_{B_s}$,
measured to be
$(3.337\pm 0.033)\times 10^{-13}$ GeV and $(11.69 \pm 0.08)\times
10^{-12}$ GeV \cite{CDF}, respectively, we will use their central values
as the inputs to constrain ${\cal N}_{d(s)b}$.
For $\Delta a_{\ell}$, we will concentrate on the muon one with $\ell=\mu$.
 The difference between the
experimental value and the SM prediction
on the muon $g-2$ is given by
$\Delta
a_{\mu}=a^{\rm exp}_{\mu}-a^{\rm SM}_{\mu}=(22\pm 10)\times
10^{-10}$ \cite{PDG06}. We will take the upper limit to bound the
free parameter ${\cal N}_{\ell\ell}$. For simplicity, we set
$\Lambda_{\UP}=1$ TeV, $1< d_{\UP}<2$, $Re{\cal N}_{qb}\sim Im{\cal
N}_{qb}$ and $Re{\cal N}_{\ell\ell}\sim Im{\cal N}_{\ell\ell}$.
To see the effects of the scalar unparticle
on  the muon $g-2$ and $B_{q}-\bar B_{q}$
mixings,
 we first show the results
 in Fig.~\ref{fig:mB-g_2},
 where the solid, dotted, dashed and dash-dotted lines stand for
$d_{\UP}=1.2$, $1.4$, $1.6$ and $1.8$, respectively.
We note that $Im{\cal N}_{db}$ is treated as a free parameter due to
${\cal N}_{sb}=\sqrt{m_d/m_s} {\cal N}_{db}$. From the figures, we
find that the muon $g-2$ and $B_{q}-\bar B_{q}$ mixings  are very
sensitive to the scale dimension $d_{\UP}$. The smaller $d_{\UP}$ it
is, the stronger constraint on $Im {\cal N}_{\mu\mu(db)}$ we get.
%%%%%%%%%%%%%%%%%%%%%%%%%%%%%%%%%%%%%%%%%%%%%%%%%%%%%%%%%%%%%
%%%%%%%%%%%%%%%%%%%%%%%%%%%%%%%%%%%%%%%%%%%%%%%%%%%%%%%%%%%%
Furthermore, with the  inputs and  the allowed values of
$Im{\cal N}_{\ell}$ and $Im{\cal N}_{db}$,
the BRs for $B_{d}\to \mu^+
\mu^-$ [solid] and $B_{s} \to \mu^{+}\mu^{-}$ [dashed] and CPA for
$B_d\to \mu^+ \mu^-$ as  functions of $d_{\UP}$ are presented in
Fig.~\ref{fig:BR-CP}.
Due to the current upper limit on ${\cal B}(B_{s}\to
\mu^{+}\mu^{-})$, $d_{\UP}$ should be less than $1.66$. The flat
curves in Fig.~\ref{fig:BR-CP}(a) correspond to the SM predictions.
Amazingly, from Fig.~\ref{fig:BR-CP}(b) we see that an unusual
direct CPA of $O(10\%)$ is generated in $B_{d}\to \mu^{+} \mu^{-}$.
Besides the necessary weak CP violating phase $\beta$ existed in the
SM, the result mainly depends on the CP conserving phase carried by
the unparticle in its propagator. This is a unique phenomenon since
it is supposed to be vanishing small even some new CP violating
phases are introduced.
In other words, if a signal of the CPA in
$B_{d}\to \mu^{+} \mu^{-}$ is observed,
it must be the
unparticle effect. Similar results are also expected in the
dielectron and
ditau modes.
However, there is no direct CPA for $B_{s}\to
\ell^{+} \ell^{-}$ decays
due to $\beta_{s}=0$ in the SM.
%%%%%%%%%%%%%%%%%%%%%%%%%%%%%%%%%%%%%%%%%%%%%%%%%%%%%%%%%%%%%
%%%%%%%%%%%%%%%%%%%%%%%%%%%%%%%%%%%%%%%%%%%%%%%%%%%%%%%%%%%%

In summary, we have studied the flavor structures of the SM fermions
when they couple to the scale invariant stuff. In order to get
naturally suppressed FCNC effects at tree level and
more correlative
transitions among three generations, we have introduced the $\BZ$ charges that
are universal in each generation
but generation un-blind.
 The $\BZ$ charges could be regarded as the internal
degrees
%}
of freedom carried by the fermions for which
the $\BZ$-fields can distinguish
the flavor generations.
 By the dimensional transmutation, the
$\BZ$ charges are matched onto the unparticle charges when
the $\BZ$ operators
onto the unparticle operators. After the EWSB, the FCNCs are induced
by the rediagonalizations of the fermion mass matrices.
To demonstrate the FCNC
effects, we have adopted the simplest
Fritzsch ansatz for quarks.
Consequently, we have found that the FCNC
effects are associated with the square roots of the mass ratios,
{\it i.e.}, $\sqrt{m_im_j/m^2_{3}}$.
In addition, although the couplings of the FCNCs could be
complex,
there is no more new CP violating phase available
because the matrices $O_{F} {\bf Q}^{\UP} R_{F}
H^{\dagger}_{F} O^{T}_{F}$ responsible to
the FCNC effects are symmetric. Moreover, we have used $\bar
B_{q}\to \ell^{+} \ell^{-}$ decays to illustrate the influence of
unparticles. In particular, with the peculiar CP conserving phases
carried by unparticles, a unique phenomenon is generated in the
direct CPAs of $B_{d}\to \ell^{+}\ell^{-}$.
If any CP violating signal is found in these decays, it must indicate the
existence of unparticles.\\

%\begin
\noindent
{\bf Acknowledgments}

We would like to thank Prof. Ling-Fong Li for useful discussions.
This work is supported in part by the National Science Council of
R.O.C. under Grant \#s: NSC-95-2112-M-006-013-MY2 and
NSC-95-2112-M-007-059-MY3.

%\end{acknowledgments}

\end{document}